\documentclass[11pt]{article}

\usepackage[margin=1in]{geometry}
\usepackage{times}
\usepackage{amsmath,amssymb}
\usepackage{graphicx}
\usepackage{booktabs}
\usepackage{hyperref}
\usepackage[round]{natbib}
\usepackage{authblk}
\usepackage{caption}

\hypersetup{
  colorlinks=true,
  linkcolor=blue,
  citecolor=blue,
  urlcolor=blue
}

\title{\textbf{HiFiC-G: Adapting HiFiC for Hi-C Contact Matrices}}
\author[1]{Straton Andre Antonio}
\affil[1]{Faculty of Mathematics and Computer Science, Transilvania University of Bra\c{s}ov (UNITBV)}
\date{}

\begin{document}

\maketitle
\vspace{-2em}

\begin{abstract}
We study whether the loss design of High-Fidelity Generative Image Compression (HiFiC), a GAN-based neural codec originally built for natural photographs, can be adapted to preserve biologically meaningful structure in Hi-C chromatin contact maps under lossy compression. Standard image compression, including HiFiC in its original form, optimizes for human visual perception; but a Hi-C contact map is normally distributed together with its numeric matrix file (.cool/.mcool), which downstream genomic analysis tools consume directly. Aggressive compression that looks acceptable to the eye can nonetheless blur or delete loops and topologically associating domain (TAD) boundaries that these tools depend on. We modify HiFiC's distortion term with a spatially-weighted MSE that up-weights biologically salient regions (loops, TAD boundaries, stripes, compartment structure) and add an insulation-score loss term that directly penalizes loss of TAD boundary sharpness. We describe a three-phase fine-tuning strategy that adapts a pretrained HiFiC checkpoint to the Hi-C domain without catastrophic forgetting. We evaluate the resulting system, HiFiC-G, using both conventional image-quality metrics (PSNR, SSIM) and genomics-domain preservation metrics (loop/TAD/compartment/stripe preservation percentage) across two cell lines. HiFiC-G preserves local structure, meaning stripes and TAD boundaries, substantially better than the metrics alone would suggest, while long-range A/B compartment structure remains poorly preserved; we show this gap tracks genomic scale and is consistent with a specific architectural cause, the fixed-size tiling that both HiFiC-G and the original HiFiC rely on for memory efficiency.
\end{abstract}

\section{Introduction}

Hi-C sequencing produces a contact matrix describing the 3D contact frequency between genomic loci, and is a central assay of the broader effort to map genome organization at scale \citep{dekker2017}. In practice, this matrix is stored as a .cool or .mcool file and is also frequently rendered as a heatmap image for visual inspection, publication figures, and manual QC. These two artifacts serve different consumers: the numeric file feeds downstream computational tools, including loop callers, TAD callers, compartment analysis, and tools such as cooltools and HiGlass, while the image is what a person looks at directly. Standard lossy image compression is designed around the second consumer and implicitly assumes that any information a human cannot perceive is safe to discard. This assumption does not hold for Hi-C: loop signals and TAD boundaries can be visually subtle, low-contrast features that carry disproportionate biological weight, and a compressor tuned purely for perceptual similarity can degrade exactly this content while still looking visually plausible to an untrained viewer.

HiFiC \citep{mentzer2020hific} demonstrated that combining a GAN loss with a learned rate-distortion objective produces reconstructions that are visually preferred to non-generative codecs at a fraction of the bitrate, for natural photographs. Our question is whether the same generative-compression machinery can be repurposed for Hi-C data if the notion of ``what must be preserved'' is redefined around genomic structure rather than natural-image statistics. We refer to our adapted system as HiFiC-G.

The main contributions of this work are:
\begin{itemize}
\item an adaptation of HiFiC, a generative image-compression architecture, to Hi-C contact matrices, including a data pipeline, staged fine-tuning procedure, and distribution-bias correction for the domain shift from natural images;
\item Hi-C-specific training objectives, namely a spatially-weighted distortion loss, an insulation-score loss targeting TAD boundary sharpness, and a compartment-pattern loss targeting long-range checkerboard structure, that redefine what ``must be preserved'' around genomic rather than perceptual salience;
\item a biological evaluation framework combining conventional image metrics with structure-preservation metrics from two independently implemented detectors, used together to bound preservation rather than relying on either alone;
\item an analysis identifying fixed-size tiling, not the loss design, as the limiting factor for long-range chromatin structure preservation under this class of architecture.
\end{itemize}

This document describes the system's design, its evaluation to date, and the limitations that bound how that evaluation should be read.

\section{Related Work}

Learned image compression is generally framed around Shannon's rate-distortion trade-off, optimizing an autoencoder and an entropy model jointly \citep{balle2018,minnen2018}. GAN-based \citep{goodfellow2014} generative compression \citep{agustsson2019,mentzer2020hific} additionally penalizes the divergence between the distribution of reconstructions and real images, formalized as a rate-distortion-perception trade-off \citep{blau2019}. HiFiC is, to our knowledge, the first such system evaluated at high resolution with a rigorous user study, and is the codebase this work builds on directly.

Separately, the Hi-C genomics literature has established standard tools for extracting biological structure from contact matrices: loop calling (e.g. HiCCUPS, \citealp{rao2014}), TAD boundary calling (e.g. TopDom, \citealp{shin2016}; Arrowhead), and reproducibility scoring between two contact matrices (HiCRep / stratum-adjusted correlation, \citealp{yang2017}). These tools were not designed with compression in mind, but they define the ground truth this work must be evaluated against: a compression method's success should be measured by whether these standard analyses give the same answer on the reconstruction as on the original, not only by pixel-level fidelity.

We are not aware of prior work that adapts a generative image-compression GAN specifically to preserve loop/TAD structure in Hi-C matrices under a domain-informed distortion loss; this is the gap this project addresses.

\section{Method}

\subsection{Background: HiFiC's Objective}

HiFiC trains an encoder $E$, generator (decoder) $G$, entropy/hyperprior model $P$, and a conditional discriminator $D$, optimizing
\begin{equation}
L_{EGP} = \mathbb{E}_{x \sim p_X}\left[ \lambda r(y) + k_M \, \mathrm{MSE}(x, x') + k_P \, d_P(x, x') - \beta \log D(x', y) \right]
\end{equation}
where $r(y)$ is the estimated bitrate of the quantized latent $y$, $d_P$ is an LPIPS perceptual distortion \citep{zhang2018lpips}, and $D$ is trained adversarially and conditioned on $y$. $\lambda$, $k_M$, $k_P$, $\beta$ are hyper-parameters controlling the trade-off between rate, distortion, and perceptual realism.

\subsection{Weighted Distortion for Biological Salience}

Our first modification replaces the uniform MSE term with a spatially-weighted variant:
\begin{equation}
\mathrm{MSE}_w(x, x') = \mathrm{mean}\left( W \odot (x - x')^2 \right)
\end{equation}
where $W$ is a per-pixel weight map computed from the input Hi-C tile prior to training. $W$ is built by detecting several structural feature classes on the tile (loop peaks, TAD boundaries, compartment (plaid A/B) structure, corner peaks, horizontal and vertical stripes) and assigning each class a multiplicative weight (in our current implementation: loops $\times 3.5$, corner peaks $\times 3.0$, stripes $\times 2.5$, TAD boundaries $\times 2.5$, compartment structure $\times 1.8$, plaid pattern $\times 1.5$), combined with an exponential decay term that keeps near-diagonal bins weighted highly regardless of feature detection, since contact frequency near the diagonal is always analytically important. Void/background regions are down-weighted to $\times 0.2$. The combined map is normalized to $[0.1, 1.0]$ before use, so that even the least important regions still receive some reconstruction pressure and training remains numerically stable. Only the MSE term is weighted; the LPIPS perceptual term is deliberately left unweighted and continues to run on the full tile, preserving global visual coherence.

\subsection{Insulation-Score Loss}

Weighted MSE penalizes large pixel-wise deviations at annotated boundary pixels, but a model can still flatten the local contrast profile around a TAD boundary (the visual signature of a boundary being ``blurred out'') without necessarily producing a large per-pixel error if the surrounding region is otherwise reconstructed well. To catch this failure mode directly, we add an auxiliary loss computed on the insulation-score profile: for each bin along the diagonal, we average contact frequency in a diamond-shaped window (10 bins $\approx$ 250 kb at 25 kb resolution in our current setup) and compare the resulting 1-D profile between original and reconstruction via MSE. A flattened profile in the reconstruction, even if the raw pixels are individually close to the original, produces a non-trivial insulation loss, giving the model a direct gradient signal to preserve boundary sharpness rather than only average intensity.

The full training objective becomes the original HiFiC loss plus $k_I \cdot L_{\mathrm{ins}}$, with $k_I$ set as a small fixed weight (currently 0.1) relative to the main distortion terms.

\subsection{Staged Domain Adaptation}

Rather than training from scratch, HiFiC-G is fine-tuned (in PyTorch, \citealp{paszke2019}) from a pretrained natural-image HiFiC checkpoint (\texttt{hific\_low}) in three phases, to avoid catastrophic forgetting of the pretrained rate-distortion behavior when exposed to Hi-C tile statistics, which differ substantially from natural photographs:
\begin{itemize}
\item \textbf{Phase 1}: freeze the network except the encoder's first convolution and the generator's last convolution (the layers that directly touch the input/output domain). This lets the network adapt to Hi-C input/output statistics without disturbing the learned internal representation.
\item \textbf{Phase 2}: unfreeze the last generator residual blocks and the hyperprior; activate the weighted-MSE and insulation losses. This is where the model begins learning which regions to preserve, now that it can represent Hi-C tiles at all.
\item \textbf{Phase 3} (optional): unfreeze the full network at a much reduced learning rate, for final refinement if Phase 2 bitrate remains too high.
\end{itemize}
Each phase uses a reduced learning rate relative to the previous one, and Phase 2/3 introduce a distillation term anchoring the fine-tuned model back toward the original pretrained checkpoint's behavior, further guarding against destabilization.

\subsection{Two Structure Detectors: Statistical and Heuristic}

The structural weight map described in Section 3.2 requires detecting loops, TAD boundaries, and A/B compartments in the input tile before training, and the same detection step is used again at evaluation time to score preservation. We implement two detectors for this, deliberately, rather than treating one as a replacement for the other. The heuristic detector uses a single global observed/expected threshold per pixel with no local background comparison or significance test, producing broad, permissive masks that tolerate the small spatial shifts and boundary smoothing tiled reconstruction introduces. The statistical detector instead follows field-standard definitions directly: TAD boundaries via the diamond-window insulation score \citep{crane2015}, A/B compartments via the leading eigenvector of the Pearson correlation matrix of the observed/expected signal \citep{lieberman2009}, and loops via a donut-kernel local background estimate with a Poisson significance test on raw contact counts, following the statistical core of the HiCCUPS method \citep{rao2014,durand2016}, implemented directly rather than via the cooltools package, which could not be built in this project's environment. Because it applies a strict, locally-normalized test rather than a fixed threshold, the statistical detector is more sensitive to exactly the kind of boundary blurring and small spatial shift that tiled reconstruction introduces (Section 3.6): a structure that survives compression only in an approximate, slightly shifted form can fail this test while still registering as preserved under the heuristic's more tolerant threshold.

A grid search over the heuristic's thresholds, scored by Jaccard overlap against the statistical detector's masks on real chromosome data, did not converge to an interior optimum: the best setting reached only 0.29--0.36 combined agreement, with loop agreement never exceeding 0.04.

This confirms the two detectors measure fundamentally different criteria, a fixed area threshold versus a locally-normalized statistical test, rather than one being a corrected version of the other, which is why we report both instead of collapsing them into one number. Section 4.2 reports the statistical detector as the primary result, since it follows established field definitions most closely; Section 4.3 reports the heuristic detector on the same data as a second, more tolerant measurement. Where the two disagree sharply, as they do for compartments (Section 5.1), the disagreement itself is evidence about where tiled evaluation breaks down, bounded on one side by the strict detector's sensitivity to reconstruction artifacts and on the other by the permissive detector's tolerance for them.

\subsection{Fixed-Size Tiling}

HiFiC-G processes the contact matrix as overlapping 256$\times$256-bin tiles (50\% stride), for the same reason the original HiFiC trains on fixed-size image crops rather than whole images: memory. A dense $N \times N$ matrix scales as $O(N^2)$, and training must additionally hold every intermediate activation for backpropagation; at 25kb resolution this scales to tens of gigabytes of GPU memory for a full chromosome, well beyond what a single consumer GPU provides. Tiling makes memory cost constant regardless of chromosome size. We verified directly in code that the weight map of Section 3.2 correctly reaches every tile, so any gap in downstream compartment preservation is not a wiring defect but a consequence of what a fixed-size tile can and cannot contain, discussed in Section 5.1.

\subsection{Compartment-Context Input Channels}

Every tile is currently encoded as three identical copies of the same local data (a legacy of warm-starting from a natural-image checkpoint expecting RGB input); channels 2 and 3 carry no information channel 1 does not already have. To address the tiling limitation of Section 3.6 without changing the network architecture, we replace these redundant channels with pre-computed long-range context: channel 2 holds the tile's row bins' compartment-identity value (the same whole-chromosome eigenvector track used in Section 3.5, broadcast across each row), and channel 3 holds the column bins' identity, broadcast down each column. This requires no architectural change, since the encoder's first convolution reads the channel count generically and does not assume the input is RGB, and fits directly into the existing Phase 1 fine-tuning stage, whose purpose is exactly to let the input-facing layers adapt to new channel statistics.

The intent is to change what the network is being asked to do: rather than inferring compartment identity from a local window that structurally cannot contain the deciding information (Section 3.6), the network is told the correct identity directly, computed once per chromosome from the full contact map, and only has to learn to produce a reconstruction consistent with it, a substantially easier, well-posed conditioning task.

This modification has been tested but not resolved. A checkpoint given the new channels without retraining reconstructs far worse than the original three-duplicate-channel scheme, as expected: the pretrained network has no prior understanding of what a broadcast compartment-identity channel means, so it cannot yet use the information usefully. Limited fine-tuning improves this somewhat, moving reconstruction quality off its untrained floor, but it remains far below the original scheme's baseline quality even after training. Section 5.2 discusses the likely cause and what would be needed to resolve this.

\subsection{Additional Auxiliary Loss: Compartment Pattern}

Mirroring the insulation-score loss of Section 3.3, we add a second auxiliary term targeting the structure that loss does not cover: the long-range A/B checkerboard pattern. Rather than a differentiable eigendecomposition inside the training loop, which risks unstable gradients whenever two eigenvalues are close (common on noisy early-training reconstructions), the loss compares the far-off-diagonal row-correlation matrix of the tile between original and reconstruction directly via MSE. This matrix is what the compartment eigenvector in Section 3.5 is itself derived from, so matching it is a stronger and numerically safer target than matching the eigenvector's scalar summary. The full training objective is therefore the original HiFiC loss plus the insulation loss (Section 3.3) plus this compartment-pattern loss, each with a small fixed weight ($k_I = k_C = 0.1$) relative to the main distortion terms.

Taken together, the modifications made to the original HiFiC codebase \citep{mentzer2020hific} in this project are: (i) a Hi-C-specific data pipeline reading .cool/.mcool contact matrices via the cooler library \citep{abdennur2020}  instead of natural images; (ii) the spatially-weighted distortion loss of Section 3.2; (iii) the insulation-score loss of Section 3.3; (iv) the compartment-pattern loss of this section; (v) the three-phase staged fine-tuning schedule of Section 3.4, including a distribution-bias correction step that measures and removes the non-zero output floor the natural-image-pretrained generator otherwise produces for true-zero Hi-C regions; (vi) the statistical and heuristic structure detectors of Section 3.5, used for weight-map construction and for evaluation; and (vii) the compartment-context input channels of Section 3.7. All are implemented as additions to or configurations of the original architecture; no changes were made to the base encoder/generator/hyperprior/discriminator network definitions themselves.

The per-chromosome preservation metrics underlying Table~\ref{tab:table1} and Table~\ref{tab:table2} (all 24 chromosomes, both cell lines, both resolutions) are available at \url{https://github.com/StratonAndre/Hific-G}. The full training and evaluation code, along with the trained checkpoints, will be added to the same repository once the compartment-context channel experiments of Section~3.7 are finalized (Section~5.2).

\section{Experiments}

The evaluation below covers the datasets, models, and metrics currently implemented; its scope and the additional baselines and statistical treatment needed for a complete evaluation are discussed in Section 5.2.

\textbf{Data:} GM12878 \citep{rao2014} and K562 Hi-C data, evaluated at 25 kb and 100 kb resolution across the full chromosome set (chr1-22, X, Y) reported in Table 1.

\textbf{Models compared to date:} pretrained HiFiC (\texttt{hific\_low} / \texttt{hific\_med} / \texttt{hific\_hi}, unmodified), an internal no-GAN baseline sharing HiFiC-G's architecture and distortion but without the adversarial term, and several fine-tuned HiFiC-G checkpoints from successive experiment iterations.

\textbf{Metrics} currently computed per model per chromosome: compression ratio, PSNR, SSIM \citep{wang2004ssim}, and the domain-specific metrics motivating this work, namely percentage preservation of loops, TAD boundaries, compartments, and stripe features between original and reconstruction, plus a noise-reduction percentage for loops/TADs (cases where the reconstruction removes spurious signal not present in the biological ground truth).

\subsubsection*{Data provenance}
K562 data is 4DN accession 4DNFI18UHVRO (confirmed human via chromosome-size matching to GRCh38) and GM12878 data is 4DN accession 4DNFIXP4QG5B (27.4GB, 68 replicates; \citealp{sanborn2015}), rather than the deposited files' portal-level labels, which were independently verified against file content before use.

\subsection{Systematic Preservation Results (Statistical Detector)}

Table~\ref{tab:table1} reports mean preservation across the full chromosome set (chr1-22, X, Y; 24 chromosomes per cell line) at both evaluated resolutions, using the medium model configuration and the statistical detector of Section 3.5.

\begin{table}[h]
\centering
\small
\begin{tabular}{lccccccc}
\toprule
\textbf{Dataset} & \textbf{Res.} & \textbf{Compression} & \textbf{PSNR} & \textbf{SSIM} & \textbf{Loops \%} & \textbf{TAD \%} & \textbf{Compart. \%} \\
\midrule
GM12878 & 25kb  & 411.1$\times$ & 35.1 dB & 0.85 & 16.7\% & 32.2\% & 9.9\% \\
GM12878 & 100kb & 274.0$\times$ & 35.2 dB & 0.93 & 25.6\% & 46.8\% & 10.8\% \\
K562    & 25kb  & 466.0$\times$ & 35.9 dB & 0.85 & 14.1\% & 62.4\% & 4.1\% \\
K562    & 100kb & 311.6$\times$ & 35.6 dB & 0.93 & 23.7\% & 57.3\% & 5.2\% \\
\bottomrule
\end{tabular}
\caption{Mean structure preservation by dataset and resolution, statistical detector, corrected source data (n=24 chromosomes per row). Stripe preservation (79--86\% across all rows) is omitted from the table for space; see text.}
\label{tab:table1}
\end{table}

The consistent pattern across both cell lines and both resolutions is that stripes are well preserved (79--86\%), TAD boundaries moderately so (32--62\%), and loops (14--26\%) and especially A/B compartments (4--11\%, median frequently at or near 0\%) are preserved poorly. This ordering matches genomic scale almost exactly and is consistent with the tiling limitation discussed in Section 3.6: stripes and TADs are near-diagonal, local phenomena that a 256-bin tile fully contains, while compartment identity depends on correlations tens of megabases outside any tile's field of view.

\begin{figure}[h]
\centering
\includegraphics[width=\textwidth]{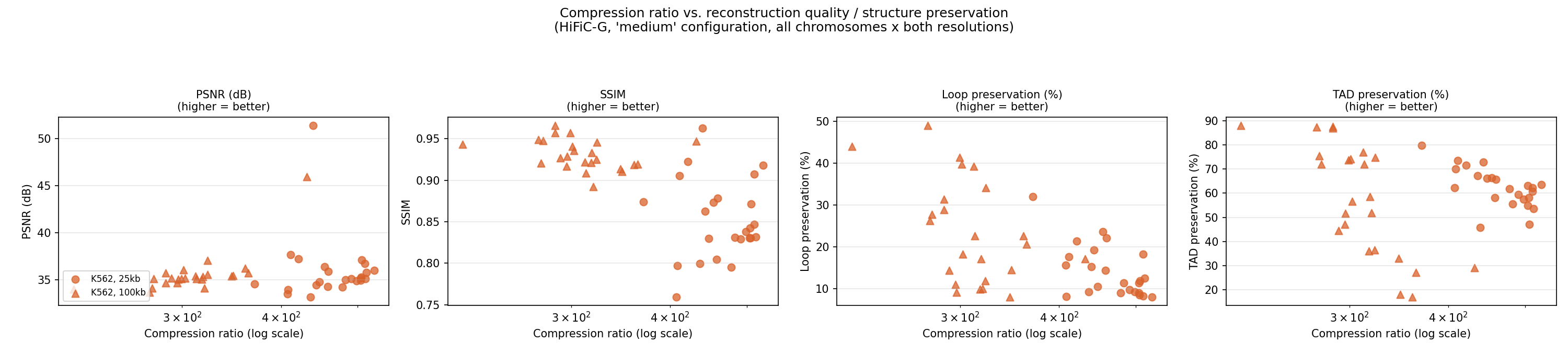}
\caption{Compression ratio vs.\ PSNR / SSIM / loop preservation / TAD preservation, all evaluated chromosomes, statistical detector (K562, both resolutions shown; mirrors Figure 4 of \citet{mentzer2020hific}).}
\label{fig:rd_curves}
\end{figure}

\begin{figure}[h]
\centering
\includegraphics[width=\textwidth]{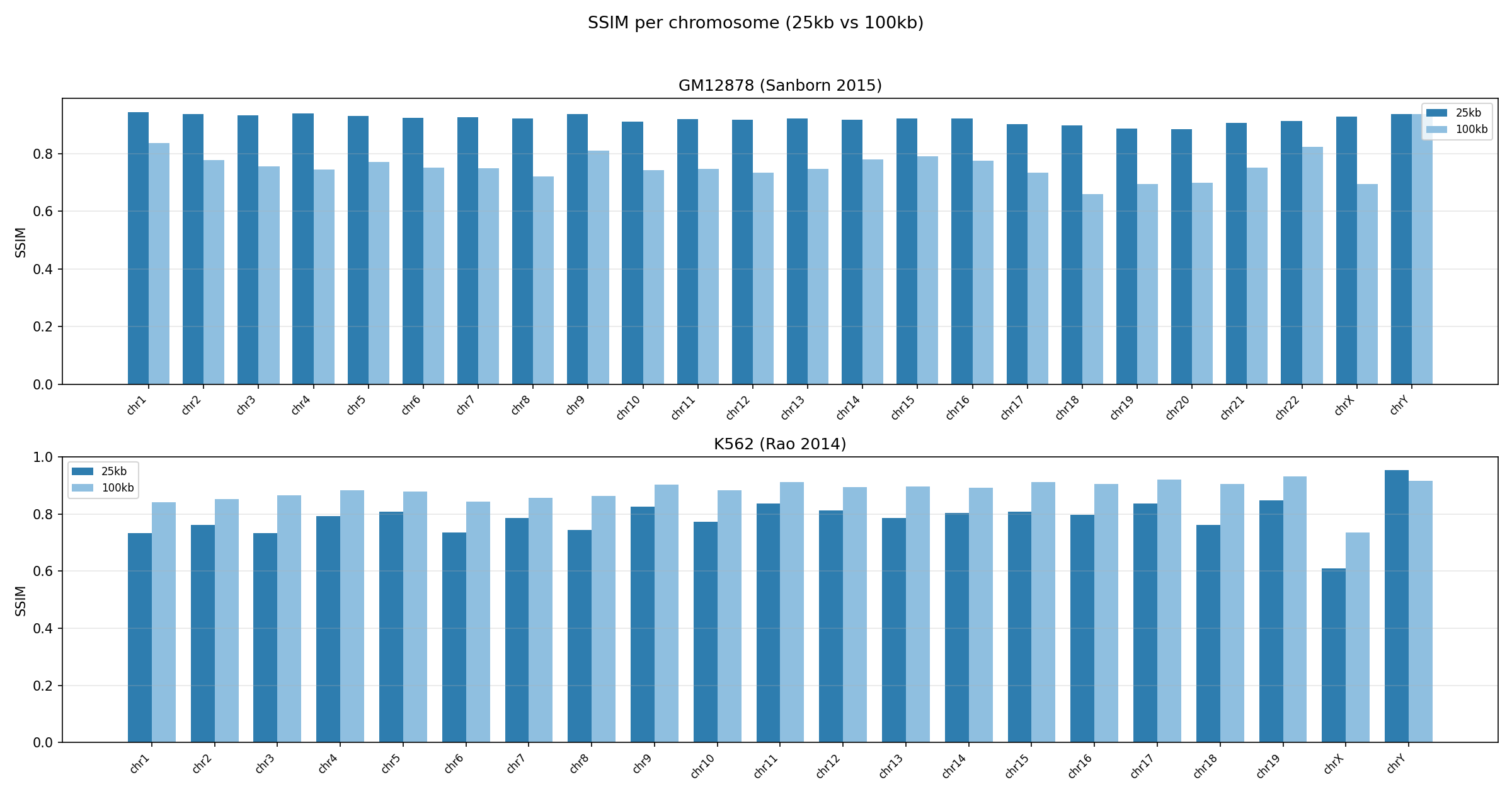}
\caption{SSIM per chromosome at 25kb and 100kb resolution, for GM12878 (top, \citealp{sanborn2015}) and K562 (bottom, \citealp{rao2014}). The two cell lines show opposite resolution effects: for GM12878, SSIM is consistently higher at 25kb than at 100kb across every chromosome, while for K562 the reverse holds, SSIM is consistently higher at 100kb. Since both datasets share the same model and evaluation pipeline, this is not an artifact of the compression method itself but reflects a difference in how each dataset's noise or signal characteristics interact with resolution, consistent with the cross-cell-line discussion in Section 4.3. Neither direction is uniformly better: the choice of resolution trades SSIM against compression ratio and structure preservation differently per cell line, and this figure is included to make that dataset-dependence visible rather than to argue for one resolution over the other.}
\label{fig:ssim_per_chrom}
\end{figure}

\subsection{Heuristic-Detector Results on the Same Data}

Table~\ref{tab:table2} reports the same comparison as Table~\ref{tab:table1} (pretrained baseline versus fine-tuned HiFiC-G), scored by the heuristic detector of Section 3.5 instead of the statistical detector, on an earlier chr22 run predating the source-data correction of Section 4.1. Because both the detector and the source data differ from Table~\ref{tab:table1}, the two tables are not directly comparable row-for-row; Table~\ref{tab:table2} is included to show the heuristic detector's substantially higher preservation numbers on the same underlying comparison, consistent with the tolerance-to-tiling-artifacts behavior described in Section 3.5, rather than as an updated or corrected version of Table~\ref{tab:table1}.

\begin{table}[h]
\centering
\small
\begin{tabular}{lccccccc}
\toprule
\textbf{Variant} & \textbf{Compression} & \textbf{PSNR} & \textbf{SSIM} & \textbf{Loops \%} & \textbf{TAD \%} & \textbf{Compart. \%} & \textbf{Stripes \%} \\
\midrule
hific\_med (baseline) & 416.2$\times$ & 38.2 dB & 0.95 & 95.2\% & 82.6\% & 64.9\% & 98.1\% \\
HiFiC-G (fine-tuned)  & 165.7$\times$ & 28.9 dB & 0.95 & 94.3\% & 84.6\% & 71.4\% & 98.1\% \\
\bottomrule
\end{tabular}
\caption{Earlier chr22 ablation-trail result (custom heuristic detector, pre-correction source data), reproduced for comparison against Table~\ref{tab:table1}.}
\label{tab:table2}
\end{table}

The gap between Table~\ref{tab:table2} and Table~\ref{tab:table1}, for example compartment preservation of 64.9--71.4\% versus 4.1--11.0\%, has two contributing sources, and we cannot fully separate them with the data on hand. First, the two tables use different detectors: the heuristic's single, unnormalized O/E threshold produces broad masks that overlap between original and reconstruction under a more permissive criterion than the statistical detector's locally-normalized significance test (Section 3.5), so some of the gap reflects detector sensitivity rather than a difference in what was actually reconstructed. Second, Table~\ref{tab:table2}'s underlying run used chr22 data predating the source-data correction of Section 4.1 and has not been independently re-verified against it. Given both factors, we do not treat either table's compartment number as the preservation rate for this system; we report both as bounds and flag compartment preservation as unresolved pending the whole-chromosome, non-tiled evaluation described in Section 5.2. Loop, TAD, and stripe preservation, where the two detectors agree more closely, are less affected by this ambiguity.

\subsection{Cross-Cell-Line TAD Preservation Difference}

Table~\ref{tab:table1} shows K562 TAD preservation (57.3--62.4\%) meaningfully higher than GM12878's (32.2--46.8\%) at both resolutions. Before the correction in Section 4.1, this specific comparison could not have been made validly at all: the file then used for K562 was actually mouse data, so any GM12878-vs-K562 difference observed at that time would have conflated a real cell-line difference with an undiscovered species difference. With both datasets now confirmed human, the residual TAD gap is a genuine, open question rather than an artifact: it may reflect an actual biological difference in TAD boundary sharpness between a lymphoblastoid line (GM12878) and a myeloid leukemia line (K562), or a detector-sensitivity interaction with each cell line's specific contact-map statistics (e.g. sequencing depth or noise characteristics differing between the two source experiments). We have not yet distinguished between these explanations and do not draw a biological conclusion from this gap here; resolving it is listed as future work in Section 5.2.

\begin{figure}[h]
\centering
\includegraphics[width=\textwidth]{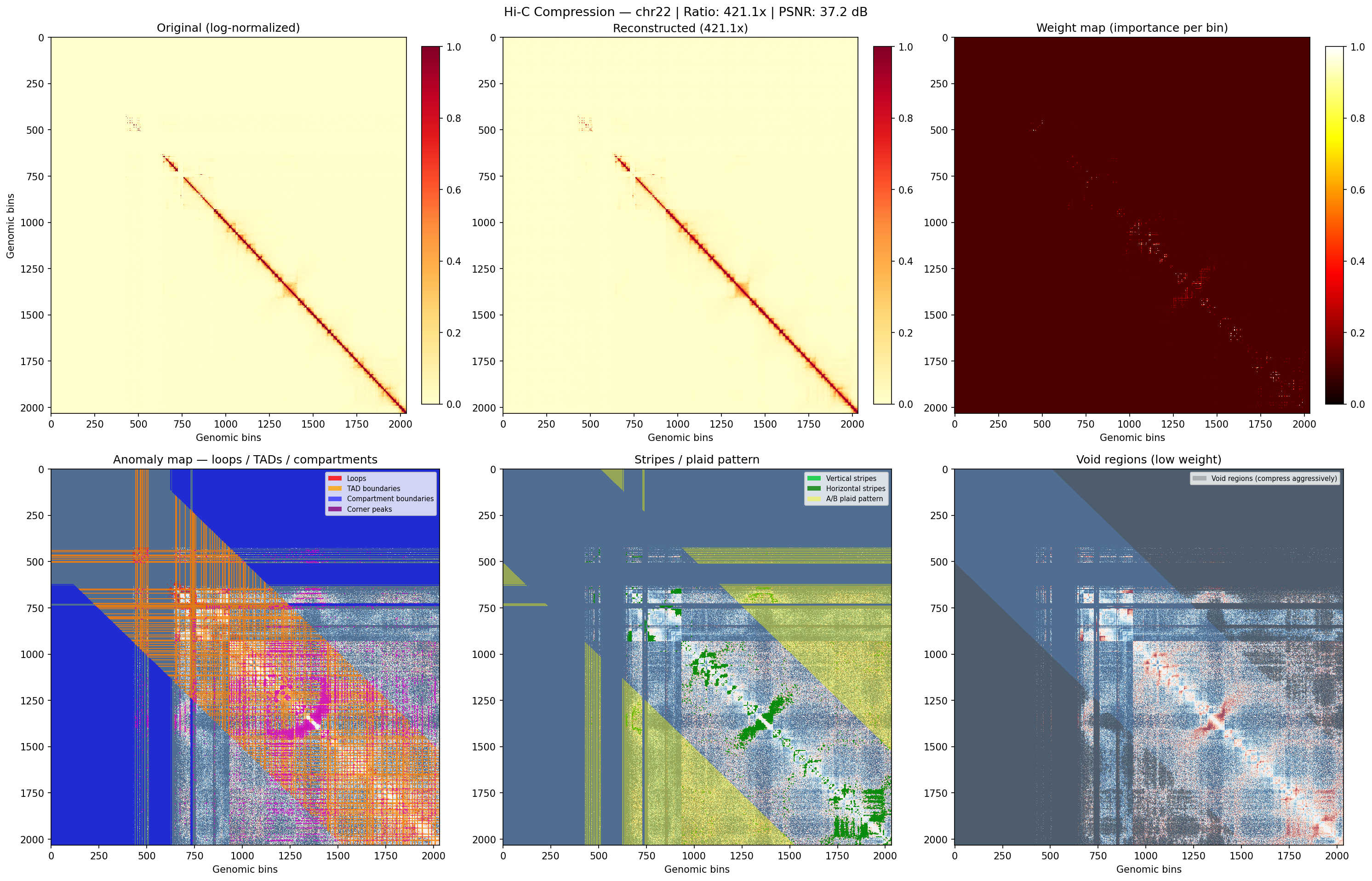}
\caption{Example per-chromosome output panel (K562, chr22, 25kb, statistical detector): original matrix, reconstruction, weight map, and detected-structure overlays, generated automatically for every evaluated chromosome. See Section 4.4 for how this figure should and should not be used.}
\label{fig:example_panel}
\end{figure}

\subsection{Note on Result Visualizations}

Diagnostic panels such as Figure~\ref{fig:example_panel} (original matrix, reconstruction, weight map, and detected-structure overlays) are generated automatically for every evaluated chromosome, intended for the researcher's own visual inspection during development: they are copies of the same kind of figure a researcher would look at when spot-checking a run, not a distinct evidentiary artifact. They are illustrative, not evidentiary: a reconstruction can look visually plausible (as reflected in a healthy PSNR/SSIM) while still failing rigorous structural preservation tests, exactly the risk named in the Broader Impact section below. No claim in this document about whether a given reconstruction is biologically faithful is based on visual inspection of a panel like this; every preservation figure in Table~\ref{tab:table1} and Table~\ref{tab:table2} comes from independently re-running the structure detector on the reconstruction and comparing to the original, never from looking at the image alone.

\section{Discussion}

\subsection{Why Compartment Preservation Remains Difficult}

The consistent finding across Sections 3.6, 4.2, and 4.3 is that A/B compartment structure is preserved substantially worse than loops, TAD boundaries, or stripes, and that this gap is not simply a scale issue in the way pixel-level distortion is: it reflects a mismatch between what the architecture can see and what the structure requires to be identified at all. Loops, TAD boundaries, and stripes are near-diagonal phenomena: their defining signal falls within a few hundred kilobases of the diagonal, comfortably inside a single 256-bin (6.4 Mb) training tile. Compartment identity is not local in this sense: it is defined by correlated contact patterns between loci tens of megabases apart, patterns that a fixed-size tile cannot represent because the correlated partner is, by construction, often outside the tile entirely. A model trained tile-by-tile can therefore learn to preserve local structure well while having no consistent way to learn what determines compartment identity, independent of how well-designed its loss function is.

This also explains why the two structure detectors disagree so much more sharply on compartments than on any other class (Section 4.3): a reconstruction can retain enough compartment-consistent signal to keep PSNR and SSIM high, and to satisfy a permissive threshold-based detector, while still failing a stricter, locally-normalized statistical test whose sensitivity is tuned to exactly the kind of boundary blurring that tiled reconstruction introduces. Read together, the two detector outputs bound the problem rather than resolve it: the true preservation rate is neither the strict number nor the permissive one, and distinguishing between ``the structure was destroyed'' and ``the structure survived in a form the strict detector cannot certify'' requires evaluating on whole, untiled chromosomes, which is outside the compute available to this project (Section 5.2). We treat this as the central open scientific question raised by this work, rather than as a deficiency to be smoothed over: architectures built around fixed-size local windows have a structural reason to struggle with genomic features defined at long range, and Section 3.7's compartment-context channels are a direct, if not yet fully evaluated, response to that finding.

\subsection{Limitations and Future Work}

This work has several limitations that bound how its results should be read, and that define the next steps for the project.

\textbf{Compartment-context input channels do not yet work.} The compartment-context channels of Section 3.7 have been implemented and mechanism-checked: the modified data pipeline and model produce correct, well-formed tensors and train without error. Reconstruction quality, however, has not recovered to a usable level in testing to date. A checkpoint given the new channels without retraining reconstructs far worse than the original scheme, confirming the network depends on its input channel meaning rather than ignoring the extra channels; after limited fine-tuning (on the order of an hour of training on this project's hardware, well short of the full staged schedule), reconstruction quality improves modestly but remains far below the original three-duplicate-channel scheme's baseline (PSNR in the 9--16 dB range against a baseline of 35--38 dB). The most likely explanation is capacity rather than a flawed idea: Phase 1 unfreezes only a small boundary layer, and Phase 2 only the generator tail and hyperprior, leaving most of the pretrained encoder frozen and unable to adapt to the new channels' different statistical structure. A brief attempt at Phase 3 (unfreezing the full network) was stopped well before completion due to its much higher per-step cost on this project's hardware, so whether full-network fine-tuning resolves this remains untested. This is treated as an open, ongoing experiment rather than a resolved result, and is not used to support any claim about compartment preservation in Section 4.

\textbf{Detection depends on two internal, non-standard detectors.} Both the statistical and heuristic detectors of Section 3.5 were implemented directly in NumPy \citep{harris2020} and SciPy \citep{virtanen2020} rather than via an established package such as cooltools, since the compiled cooltools distribution could not be built in this project's environment. The underlying algorithms follow published definitions, but the implementations have not been cross-validated against a reference caller such as HiCCUPS or TopDom. Reporting precision/recall/F1 of calls against such a caller, rather than raw preservation percentages from either internal detector, would substantially strengthen the preservation results in Section 4.

\textbf{Compartment preservation cannot yet be attributed to a single cause.} As discussed in Section 5.1, resolving how much of the statistical/heuristic detector gap reflects a tiling artifact versus a genuine reconstruction failure requires evaluating on whole, non-tiled chromosomes, or with a tiling scheme that preserves long-range context. This is not currently practical on this project's hardware and is the single most important piece of unfinished evaluation in this work.

\textbf{Limited external baselines.} Section 4 compares HiFiC-G against unmodified pretrained HiFiC and an internal no-GAN ablation, but not against a generic lossless Hi-C storage baseline, a generic image codec (e.g. JPEG2000) at matched bits-per-pixel, or any Hi-C-specific compression method in the existing literature, should one exist; a literature search for such a method has not yet been completed. Without these, the absolute case for HiFiC-G's compression ratios cannot yet be made, only the relative case between HiFiC-G and unmodified HiFiC.

\textbf{Limited statistical treatment and evaluation scope.} Table~\ref{tab:table1} reports mean preservation across 24 chromosomes per cell line, but not standard deviation or a paired significance test, and evaluation to date covers two cell lines (GM12878, K562) rather than the wider dataset panel available (Section 4.1). Reporting rate-distortion-preservation curves across a range of bitrates, mirroring Figure 4 of \citet{mentzer2020hific}, rather than single operating points, would also make the comparison to the original HiFiC more direct. A small human evaluation with genomics-trained raters, a domain-specific analogue of HiFiC's 2AFC user study, would add a form of validation this work does not currently have.

\textbf{No isolated ablation of the individual loss terms.} The results in Section 4 reflect the combined weighted-MSE and insulation-score losses together; a four-way ablation (vanilla HiFiC / +weighted-MSE only / +insulation-loss only / both combined) would isolate each term's individual contribution and has not yet been run.

\section{Conclusion}

We have presented HiFiC-G, an adaptation of HiFiC to Hi-C contact matrices that replaces its distortion term with a biologically weighted alternative and adds two auxiliary losses targeting TAD boundary sharpness and long-range compartment pattern specifically, together with a staged fine-tuning procedure that adapts a pretrained natural-image checkpoint to this new domain. Evaluated across two cell lines and two resolutions against unmodified HiFiC, HiFiC-G preserves stripe and TAD structure at levels that track genomic scale, and preserves A/B compartment structure poorly under a strict, field-standard detector, a result we attribute to the architecture's fixed-size tiling rather than to the loss design, since compartment identity is defined at a genomic distance the tiling scheme cannot represent (Section 5.1). Section 5.2 details the further evaluation needed to state this more precisely, including cross-validation against established structure callers and non-tiled compartment evaluation. Our results suggest that adapting learned image compression to Hi-C data requires optimization objectives that reflect genomic structure rather than visual appearance: metrics tuned for human perception do not reliably track biological fidelity, and the two can diverge sharply for structure defined at long genomic range. More broadly, our findings indicate that preserving long-range chromatin organization is a fundamental challenge for fixed-tile architectures specifically, rather than an artifact of any particular loss design, motivating context-aware compression models as a direction worth pursuing, though our own first attempt at one, the compartment-context channels of Section 3.7, has not yet succeeded and remains an open experiment (Section 5.2).

\section*{Broader Impact}

As with the original HiFiC, a generative compressor can in principle synthesize plausible-looking structure that was not present in the original data. This risk is more consequential in a genomics context than in natural photography: a hallucinated loop or a smoothed-over TAD boundary could mislead downstream biological interpretation if the compressed reconstruction were mistaken for, or used in place of, the original matrix. Any released model or tool should be clearly labeled as a lossy compression artifact and should not be treated as a substitute for the original .cool/.mcool file in contexts where exact analytical fidelity is required (e.g. clinical or diagnostic use). This caveat should be stated explicitly in any published version of this work.

\section*{Acknowledgments}

This work was carried out as part of a Bachelor's thesis at the Faculty of Mathematics and Computer Science, Transilvania University of Bra\c{s}ov (UNITBV). The research itself was conducted independently, without dedicated Hi-C lab infrastructure, compute beyond a personal laptop GPU, or review by Hi-C domain experts prior to release.

\bibliographystyle{plainnat}
\bibliography{references}

\end{document}